\documentclass[12pt]{article}
\pdfoutput=1
\usepackage{titling}
\usepackage{amsmath}
\usepackage{slashed}
\usepackage{amssymb}
\usepackage{epsfig}
\usepackage{graphicx}
\usepackage{multirow}
\usepackage{color}
\usepackage[normal]{subfigure}
\usepackage{rotating}
\usepackage{hyperref}
\usepackage[margin=0.9in]{geometry}
\usepackage[table]{xcolor}
\usepackage{enumitem}
\usepackage[utf8x]{inputenc}
\usepackage[compress,numbers,sort]{natbib}
\usepackage{authblk}
\usepackage{colortbl}
\usepackage{pdflscape}
\usepackage{color}
\usepackage{mathtools}

\definecolor{nicered}{rgb}{0.7,0.1,0.1}
\definecolor{nicegreen}{rgb}{0.1,0.5,0.1}
\definecolor{red}{rgb}{1.0, 0, 0}
\hypersetup{colorlinks,citecolor= nicegreen,linkcolor= nicered}

\def\eq#1{{Eq.~(\ref{#1})}}
\def\eqs#1#2{{Eqs.~(\ref{#1})--(\ref{#2})}}
\def\fig#1{{Fig.~\ref{#1}}}

\def\Table#1{{Table~\ref{#1}}}
\def\Tables#1#2{{Tables~\ref{#1}--\ref{#2}}}
\def\sect#1{{Sect.~\ref{#1}}}

\def\app#1{{Appendix~\ref{#1}}}

\def\abs#1{\left| #1\right|}

\def\Re{\mbox{Re}\,}
\def\Tr{\mbox{Tr}\,}

\definecolor{LightCyan}{rgb}{0.88,1,1}
\definecolor{piggypink}{rgb}{0.99, 0.87, 0.9}
\definecolor{applegreen}{rgb}{0.55, 0.71, 0.0}
\definecolor{darkpastelgreen}{rgb}{0.01, 0.75, 0.24}
\definecolor{green-yellow}{rgb}{0.68, 1.0, 0.18}

\newcommand{\beq}{\begin{equation}}
\newcommand{\eeq}{\end{equation}}
\newcommand{\bea}{\begin{eqnarray}}
\newcommand{\eea}{\end{eqnarray}}

\newcommand{\published}[1]{%
\gdef\puB{#1}}
\newcommand{\puB}{}

\title{\bf{\huge Massive vectors and loop observables: the $g-2$ case}}

\author[1]{\large Carla Biggio\thanks{carla.biggio@ge.infn.it}}
\author[2]{\large Marzia Bordone\thanks{mbordone@physik.uzh.ch}}
\author[1]{\large Luca Di Luzio\thanks{luca.di.luzio@ge.infn.it}}
\author[1]{\large Giovanni Ridolfi\thanks{giovanni.ridolfi@ge.infn.it}}
\affil[1]{\emph{\normalsize Dipartimento di Fisica, Universit\`a di Genova and INFN, Sezione di Genova, \newline
via Dodecaneso 33, 16146 Genova, Italy}}
\affil[2]{\emph{\normalsize Physik-Institut, Universit\"at Z\"urich, CH-8057 Z\"urich, Switzerland}}

\date{}

\published{\flushright ZU-TH-23/16\vskip1cm}

\begin{document}

\maketitle

\begin{abstract}
\normalsize
We discuss the use of massive vectors for the interpretation of some
recent experimental anomalies, with special attention to the muon
$g-2$. We restrict our discussion to the case where the massive vector
is embedded into a spontaneously broken gauge symmetry, so
that the predictions are not affected by the choice of an arbitrary energy
cut-off. Extended gauge symmetries, however, typically impose strong
constraints on the mass of the new vector boson and for the muon 
$g-2$ they basically rule out, barring the case of abelian gauge
extensions, the explanation of the discrepancy in terms of a single
vector extension of the standard model. We finally comment on the use
of massive vectors for $B$-meson decay and di-photon anomalies.

\end{abstract}

\clearpage

\tableofcontents

\clearpage

\section{Introduction}

In the recent years there has been quite a lot of interest for the emergence of a few $3$-$4\sigma$ 
experimental anomalies 
in particle physics. Among those, the most relevant are the longstanding one of the 
anomalous magnetic moment of the muon, $(g-2)_\mu$,~\cite{Bennett:2006fi} 
(see Ref.~\cite{Jegerlehner:2009ry} for a review) and a collection of anomalies 
in semileptonic $B$-meson decays~\cite{Aaij:2013qta,Aaij:2014ora,Aaij:2015yra}. 
More recently, ATLAS \cite{ATLAS-CONF-2015-081,Aaboud:2016tru} and CMS \cite{CMS:2015dxe,CMS:2016owr,Khachatryan:2016hje} 
reported a hint of a di-photon resonance with mass in the vicinity 
of $750$ GeV in the first LHC data collected at 13 TeV collision 
energies.\footnote{New 2016 LHC data at 13 TeV have not confirmed the excess \cite{ATLAS:2016eeo,CMS:2016crm}.} 
None of them is conclusive at the moment, and require further scrutiny both from the experimental 
and the theoretical point of view; it is nevertheless tantalizing to try to interpret them 
within new physics frameworks beyond the standard model (SM). 
This has triggered a large amount of works, ranging from full-fledged
theoretical constructions, like for example supersymmetry, up to simplified 1-particle extensions of the SM. 
In the latter case, one simply adds a new irreducible representation (irrep) on top of the SM field content, 
with spin quantum number $0$, $1/2$, $1$, etc. While the case of a new scalar or fermion irrep is conceptually 
straightforward, being the SM extension automatically renormalizable and well-behaved in the ultraviolet (UV), 
the one of a generic Lorentz vector is less obvious and will be the subject of the present paper.  

The two main challenges that one faces when extending the SM with a vector irrep are the following: 
$i)$ depending on the UV completion, the theory might not be renormalizable, 
thus reducing the degree of predictivity for the observables whose anomaly one is willing to explain and 
$ii)$ regardless of the renormalizability issue, the 1-particle
extensions hypothesis is possibly violated in explicit constructions,
which require several new particles at the same energy scale.

Concerning the first point, massive vectors typically arise either as composite states resulting from a new strongly-coupled dynamics 
(for example the $\rho$ meson in QCD) or as extra gauge bosons associated with a spontaneously broken 
gauge extension of the SM.   
The difference between these two possibilities is substantial, the most dramatic being renormalizability.  
Though there is nothing wrong in contemplating a non-renormalizable theory within an effective field theory (EFT) approach, 
we will focus on UV-complete, weakly-coupled models which provide a more predictive framework for dealing with precision 
loop observables. 
As a prototypical example we will mainly discuss the $(g-2)_\mu$, while commenting {\it en passant} on other anomalies.  

After a brief review of the $(g-2)_\mu$ discrepancy in \sect{gm2anomaly}, 
we discuss in \sect{EFTgm2} the most general $d \leq 4$ Lagrangian 
of a massive vector coupled to the SM, and show the divergence structure of the one-loop diagrams.  
In the particular case at hand, we will see that the culprit of the non-renormalizability 
resides in the triple vector boson vertex which has to be properly modified 
in order for the theory to be renormalizable. 
In \sect{RENgm2} we classify all possible SM gauge quantum numbers of the new vector, 
hereafter denoted by $X$, coupling to a muon and to another SM fermion 
(a general classification of the $X$ gauge quantum numbers such that it couples to SM fields at the renormalizable level 
is provided in \app{classX}). Next, by assuming that $X$ is a gauge boson of an extended SM gauge group, 
we compute for each case the finite contribution to the $(g-2)_\mu$ 
and estimate the required mass scale, $M_X$, in order to explain the discrepancy. 
Remarkably, after providing a minimal gauge embedding for each case, we find that the UV theory imposes 
strong direct and indirect constraint (e.g.~from proton decay or flavor violating processes), 
such that most of the simplified 1-particle extended models 
cannot provide an explanation of the $g-2$ discrepancy in the full 
renormalizable setup. The only exception to this rule is given by abelian gauge extensions, 
like e.g.~the case of a light dark photon or dark $Z$. 
Furthermore, another aspect emerging from the full analysis is that extra states required by the 
consistency of the gauge symmetry breaking pattern cannot be arbitrarily decoupled from $X$, 
thus typically violating the 1-particle extension hypothesis. We finally conclude in \sect{concl} by summarizing 
our findings and comment on the use of massive vectors for the $B$-meson decay and di-photon anomalies.

\section{Review of the $(g-2)_\mu$ discrepancy} 
\label{gm2anomaly} 

Known respectively with 12 and 9 digits, the anomalous magnetic moments of the electron
and the muon are among the best measured quantities in physics. While
the former is used to fix the value of the fine structure constant
$\alpha_{\rm em}$, the latter constitutes a good observable where to
look for new physics. 

The world average of the measured $a_\mu \equiv (g_\mu -2)/2$, dominated by
the result obtained by E821 at Brookhaven~\cite{Bennett:2006fi}, is given
by~\cite{Jegerlehner:2009ry}
\begin{equation}
a_\mu^{\rm{exp}}
=116592080 (63) \cdot 10^{-11} \, .
\end{equation}
In the SM $a_\mu$ arises at one loop and, due to the great precision
of this measurement, higher order corrections must be taken into
account. The SM contribution can be divided into three categories:
$i$) QED contributions, consisting of loops involving only leptons and
photons, $ii$) electroweak contributions, involving leptons, $W$, $Z$ and Higgs
bosons and $iii$) hadronic contributions, with hadronic resonances
circulating in the loops. The QED contribution has been calculated up to five
loops and the electroweak one up to two loops, which is enough for
the current experimental precision. On the other hand, the largest
error on the theoretical determination comes from the hadronic
contributions: in the light-by-light scattering amplitude some theoretical input
is needed in order to perform the calculation, while in the vacuum
polarization diagrams some dispersion relations are extracted from experiments,
either from $e^+e^-$ scattering or from $\tau$ decay. Depending on
these different inputs, different results are obtained for the
theoretical prediction. We choose
as a reference value for the SM determination the one contained
in the review~\cite{Jegerlehner:2009ry}, while a list of other
predictions can be found for instance in Ref.~\cite{Marciano:2016yhf}:
\begin{equation}
a_\mu^{\rm{SM}} =  116591790 (65) \cdot 10^{-11} \, .
\end{equation}
If we now compare this with the measured value, we get a difference of $\Delta a_\mu
= 290(90)  \cdot 10^{-11}$ which corresponds to a discrepancy with
3.1$\sigma$ significance. By choosing different theoretical predictions
one obtains discrepancies which range from 2 to 4 $\sigma$.
New, independent measurements are expected in
the next few years by two collaborations,
E989 at Fermilab~\cite{Grange:2015fou} and E34 at JPARC~\cite{Otani:2016ixh},
and therefore the existence of a
$(g-2)_{\mu}$ anomaly will soon be confirmed or disproved; for the
moment, we stick to the available experimental result.

Even if this is not enough to claim a discovery, this discrepancy
deserves a detailed analysis. 
Basically, it can arise for two
different reasons: either $i)$ the SM prediction is not accurate, 
or $ii)$ there is some
physics beyond the SM contributing to the $(g-2)_\mu$. 

Due to the difficulties in calculating the hadronic
contributions, one could think that $i)$ is the favourite
explanation. However, if one fixes the hadronic contribution in
order to agree with $a_\mu^{\rm exp}$, deviations in the electroweak precision observables are
obtained. In particular, the Higgs mass prediction is modified and, in
order to be compatible with the measured value, large modifications of
the hadronic contribution at energies lower than 1~GeV would be
required, while this is precisely the energy region where the
experimental measurement is solid~\cite{Passera:2008jk}. Therefore, this
explanation seems to be disfavoured. 

According to case $ii)$, the discrepancy $\Delta a_\mu$ could be due to the presence
of new physics beyond the SM. Indeed, in models beyond the SM involving new particles' couplings to
muons, like for example supersymmetric models, a positive (and large) contribution
to the $(g-2)_\mu$ can be achieved quite easily. Two approaches are
therefore possible:
either one takes a model, conceived to solve another problem, and
verifies whether it can also explain this discrepancy, or tries to
classify, in a more model-independent way, which are the new
particles that can contribute to the $(g-2)_\mu$. This second
approach is the one adopted e.g.~in 
Ref.~\cite{Biggio:2014ela},
where minimal extensions of the SM with a single scalar or fermion irrep were considered (see also
Refs.~\cite{Freitas:2014pua,Queiroz:2014zfa,Chakraverty:2001yg,Cheung:2001ip}
for other analysis with a similar formulation). In
the present paper, we follow the same idea and complete the classification by adding
one massive vector to the SM field content.

\section{EFT approach to the $(g-2)_\mu$
} 
\label{EFTgm2}

A possible approach to the $(g-2)_\mu$ 
consists in adding to the SM field content a new Lorentz vector, $X^\mu$,  
without specifying the full UV completion of the theory. 
In general, the theory is non-renormalizable and one expects loop observables to be divergent. 
In this section, we discuss the $d \leq 4$ operators that can appear
in the Lagrangian of a massive vector coupled to the SM 
and analyze the divergence structure of the diagrams relevant for the $g-2$. 

\subsection{Lagrangian of a massive vector}

Before performing the actual $(g-2)_\mu$ calculation, we discuss the
Lagrangian of the new vector boson, which is assumed to
transform under a complex irrep of the SM gauge group.  As already
mentioned, we will not assume that its mass originates from a
spontaneously broken gauge symmetry.  The canonical kinetic and mass
terms of $X^{\mu}$ read\footnote{Note that by Lorentz and gauge
invariance the most general Lagrangian quadratic in $X^\mu$ is
\begin{equation}
\label{canonicalLquad}
\mathcal{L}_X^{\text{free}} 
= - \partial_\mu X^\dag_\nu \partial^\mu X^\nu 
+ \beta \, \partial_\mu X^\dag_\nu \partial^\nu X^\mu + M^2_X X_\mu^\dag X^\mu 
\, , 
\end{equation} 
where $\beta$ is a free parameter. It can be shown \cite{Weinberg:1995mt} that for $\beta =1$ the above Lagrangian describes 
the free propagation of a massive spin 1 particle. For $\beta \neq 1$ a scalar degree of freedom is included as well.}
\begin{equation}
\label{LXfree}
\mathcal{L}_X^{\text{free}} 
= - \partial_\mu X^\dag_\nu \partial^\mu X^\nu
+ \partial_\mu X^\dag_\nu \partial^\nu X^\mu 
+ M^2_X X^\dag_\mu X^\mu \, ,  
\end{equation}
with propagator 
\begin{equation} 
\label{propcan}
i \Delta^{\mu\nu} (k) = \frac{i}{k^2-M^2_X} \left( -g^{\mu\nu} + \frac{k^\mu k^\nu}{M^2_X} \right)  \, ,
\end{equation}
which is the same as the unitary gauge propagator of a massive gauge boson.

We are interested in working out the interaction term of $X^\mu$ with the photon field $A^\mu$, 
which in turn contributes to the $g-2$.
The so-called minimal coupling to electromagnetism is generated by
simply replacing ordinary derivatives in \eq{LXfree}
by covariant derivatives:
\beq
\partial_\mu X_\nu\to D_\mu X_\nu=(\partial_\mu-ieQ_XA_\mu)X_\nu,
\eeq
where $Q_X$ is the electric charge of $X$ in units of the proton charge $e$.
This is enough to make the Lagrangian of \eq{canonicalLquad} invariant
upon local gauge transformations
\beq 
X^\mu \to e^{ie Q_X \alpha(x)} X^\mu;\qquad A_\mu\to A_\mu+\partial_\mu\alpha(x)\, ,
\eeq  
where $\alpha(x)$ is the local parameter of the transformation. The resulting
coupling of the vector field $X$ is 
\beq
\label{LXem}
\mathcal{L}_X^{\text{em}} =J_\mu^{\text{em}} A^\mu - e^2 Q_X^2 \left( A_\mu A^\mu X^{\dag}_\nu X^\nu - A_\mu A^\nu X^{\dag}_\nu X^\mu \right) \, ,
\eeq
where
\beq 
J_\mu^{\text{em}} = i e Q_X \left[ 
\left( \partial_\mu X^\dag_\nu - \partial_\nu X^\dag_\mu \right) X^{\nu}
- \left( \partial_\mu X_\nu - \partial_\nu X_\mu \right) X^{\dag\nu}
\right] \, ,  
\eeq
is the conserved current of the free theory. 

On the other hand, it is easy to see that there exist extra gauge invariant terms 
not related to the minimal 
coupling. A complete classification 
of SM gauge invariant $d \leq 4$ operators involving $X$ and SM fields is given in \app{classX}, 
and the most general EFT should contain them all. 

\subsection{Divergence structure of one-loop diagrams}

The EFT described in the previous subsection is non-renormalizable because $X^\mu$ is not a gauge boson. 
This can be proved on general grounds. However, it is interesting to see how non-renormalizability manifests 
itself in the case of the $g-2$, and to study its relationship with the minimal coupling. 
To this purpose we extend the minimally coupled theory by adding a gauge invariant term proportional to (see also Ref.~\cite{Barbieri:2015yvd})
\beq 
\left( X_\mu X^\dag_\nu - X_\nu X^\dag_\mu \right) \partial^\mu A^\nu \, .
\eeq
By including also the interaction terms with the muon field $\mu$ and a generic SM fermion $f$, 
the effective Lagrangian relevant for the $(g-2)_\mu$ calculation is
\begin{align}
\label{Lintgm2}
\mathcal{L}^{g-2}_{\rm int} &= 
\overline{\mu} \left( g_V \gamma_\alpha + g_A \gamma_\alpha \gamma_5 \right) f X^\alpha + \text{h.c.} \nonumber \\
& + i e Q_X
\left[ 
\left( \partial_\mu X^\dag_\nu - \partial_\nu X^\dag_\mu \right) A^\mu X^{\nu}
- \left( \partial_\mu X_\nu - \partial_\nu X_\mu \right) A^\mu X^{\dag\nu}
+ k_{Q} \left( X_\mu X^\dag_\nu - X_\nu X^\dag_\mu \right) \partial^\mu A^\nu
\right] \nonumber \\
& + i e Q_f \overline{f} \gamma_\mu f A^\mu \, ,
\end{align}
where $g_{V,A}$ are vector and axial couplings, $Q_f$ is the electromagnetic (EM) charge of $f$ 
and $k_Q$ is a free parameter.  
 
\begin{figure}[h]
\centering
\vspace*{-1cm}
\hspace*{-1.7cm}
\includegraphics[angle=0,width=10cm]{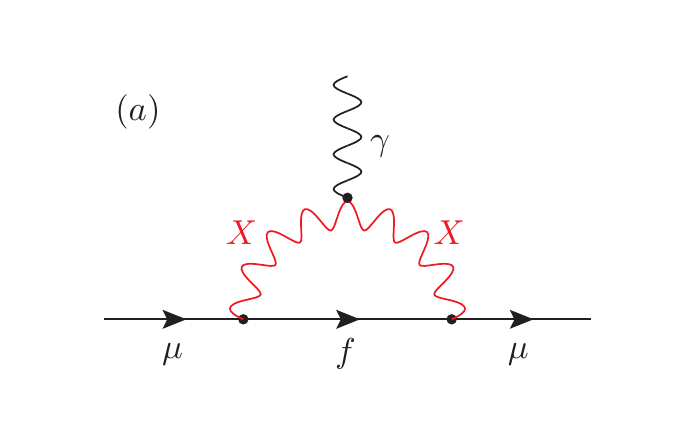}
\hspace*{-1.6cm}
\includegraphics[angle=0,width=10cm]{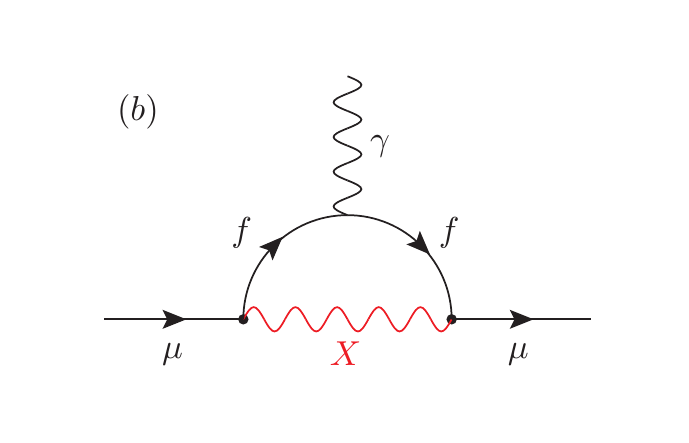} 
\hspace*{-1.8cm}
\vspace*{-1.0cm}
\caption{
\label{diagm2} One-loop diagrams contributing to the $(g-2)_\mu$. Red wiggled lines stand for the massive vector $X$, while the 
blobs in the vertices denote the interactions of $X$ with the SM fields defined in \eq{Lintgm2}.
} 
\end{figure}

The two diagrams contributing to the $(g-2)_\mu$ at the one-loop level are displayed in \fig{diagm2}.
The degree of superficial divergence of diagrams 
$(a)$ and $(b)$ is respectively $4$ and $2$. 
However, denoting by $\Lambda$ the cut-off regulator, 
an explicit calculation shows that
\begin{itemize}
\item The contribution to the $(g-2)_\mu$ from diagram $(a)$ is only logarithmically divergent, 
since the $\Lambda^4$ term vanishes when the virtuality of the external photon is set to zero, while 
the $\Lambda^2$ term goes into the renormalization of the electric charge.  
\item Diagram $(b)$ is finite.
\end{itemize}
The reduction of the degree of divergence for the 3-point function is a simple consequence of the Ward identity, 
which connects the $\mu\mu\gamma$ vertex $\Gamma^\alpha(p,q)$
to the derivative of the muon self-energy $\Sigma(p)$
in the soft-photon limit $q\to 0$ via the relation 
\beq 
\label{WI}
\Gamma^{\alpha} (p,p) = \frac{d \Sigma(p)}{d p_\alpha} \, .
\eeq 
To see this, let us Taylor expand the muon self-energy in powers of $\slashed{p}-m_\mu$ 
\beq 
\Sigma(p) = A + B (\slashed{p}-m_\mu) + \Sigma_c(p) (\slashed{p}-m_\mu) \, .
\eeq
Since $\Sigma(p)$ is linearly divergent, the first two coefficients $A$ and $B$ are respectively 
linearly and logarithmically divergent (indeed every derivative with respect to $p$ lowers the degree of divergence by one unit). 
This implies that $d \Sigma(p) / d p_\alpha$, and hence $\Gamma^{\alpha} (p,p)$ because of \eq{WI}, 
can be at most logarithmically divergent. 

By employing the Lagrangian in \eq{Lintgm2} we find the following contribution to the divergent part of the $(g-2)_\mu$:
\begin{equation}
\label{amunonren}
 \Delta a_{\mu}^{\rm div.} = \frac{Q_{X}m_{\mu}^2}{8\pi^2M_{X}^2}(k_{Q} - 1)\left[\left(g_{V}^2+g_{A}^2\right)-\frac{m_{f}}{m_{\mu}}\left(g_{V}^2-g_{A}^2\right)\right]\log{\frac{\Lambda^2}{M_{X}^2}}\, .
\end{equation} 
This result shows that the logarithmic divergence disappears in the limit $k_{Q} \to 1$.
On the other hand, the divergence persists in the minimally coupled theory ($k_{Q} = 0$). 
Also note that for $k_{Q} = 1$ and $Q_{X} =1$,  
the second line of \eq{Lintgm2} reproduces the SM triple gauge vertex $WW^{\dagger}A$, 
with the identification $X^\mu = W^{+\mu}$. We hence conclude that the choice $k_{Q} = 1$ 
is a necessary condition for renormalizability. Moreover, possible extra gauge invariant terms in \eq{Lintgm2} 
do not arise in renormalizable theories (cf.~the discussion in \app{classX}).  

Even though one could estimate the contribution of the 
massive vector to the $(g-2)_\mu$ by setting $\Lambda$ to the value of the cut-off of the EFT, 
this requires the specification of a new energy scale (e.g.~the scale of compositeness in strongly-coupled theories). 
Once an appropriate number of counterterms are fixed in terms 
of physical observables, EFTs can be fully predictive within their range of validity 
and at a given order in the coupling/energy expansion (cf.~e.g.~the case of the SM EFT \cite{Jenkins:2013zja,Jenkins:2013wua,Alonso:2013hga}). 
Nevertheless, renormalizable setups provide us with a larger degree of predictivity and in the following we will 
focus for simplicity on UV-complete, weakly coupled models.

\section{Renormalizable approach to the $(g-2)_\mu$
}
\label{RENgm2}

In this section we discuss the case in which the new vector is embedded in a 
spontaneously broken extended gauge symmetry. Hence, $k_{Q}=1$ in \eq{Lintgm2}, 
so that the vector contribution to the $(g-2)_\mu$ turns out to be finite and predicted in terms of 
a renormalizable Lagrangian. 

Before turning to the actual discussion of the gauge embeddings, we estimate the mass scale $M_X$ 
required in order to explain 
the $(g-2)_\mu$ discrepancy, regardless of its UV gauge completion. 
Since existing bounds require $M_X \gg M_W$,\footnote{The only exception is 
given by abelian gauge extensions (cf.~end of \sect{X110}). 
But in such a case $X$ is a SM gauge singlet.} it is more appropriate to employ an $SU(2)_L \otimes U(1)_Y$ invariant language. 
To this purpose,
we have classified in \app{classX} all the possible $X$-quantum numbers such that 
the new vector can couple to SM fields via $d \leq 4$ operators (cf.~\Tables{classificationVSMfermions}{classificationVSMbosons}). 
Only a subset of these operators 
are relevant for the $(g-2)_\mu$, namely all those involving a lepton
field, which are reported in \Table{classificationVgm2v2}. This Table
summarizes most of our results. It shows all the possible 
new vector's quantum numbers, together with their EM components and the $d = 4$ 
operators involving $X$, a muon and a SM fermion field. Moreover, it
contains, for each case, the sign of the contribution 
to $\Delta a_\mu$ in the approximation where the $SU(2)_L$ multiplet components have the same 
mass $M_X,$\footnote{The mass splitting between the electroweak components of an $SU(2)_L$ multiplet 
originates from a tree-level term and, for $M_X \gg M_W$,  
goes like $\Delta M_X \sim g'^2 \ 10 \ {\rm GeV} \ (1 \ {\rm TeV} / M_X)$, where $g'$ is a custodial breaking gauge coupling.  
} 
and the value of $M_X$ which is required in order to explain the
experimental discrepancy, for the reference gauge coupling $g_X=1$ ($M_X$ scales linearly with $g_X$). 
Finally, in the last column, we provide a minimal gauge embedding of the 
massive vector into an extended gauge symmetry group. What
we did not include in \Table{classificationVgm2v2} are the actual bounds on $M_X$, which are
instead discussed in detail in \sect{vectorcases}. 
In some cases a model-independent bound applies (namely without specifying the embedding), 
while in general the gauge embedding implies extra indirect constraints. 
As a matter of fact, we find that only the abelian extension can provide an explanation of the 
$(g-2)_\mu$ discrepancy, compatibly with the existing bounds.

\begin{table}[h!]
\renewcommand{\arraystretch}{1.4}
\centering
\begin{tabular}{@{} |c|c|c|c|c|c| @{}}
\hline
$X^\mu$ &  $Q_{\rm EM}$ &  $\mathcal{O}^{g-2}_X$ & $\text{sign}(\Delta a_ \mu)$ & $M_{X} [\rm GeV]$ &  Gauge embedding
\\ 
\hline
\hline
$(1,1,0)$ & 0 & 
$\overline{e}_R \gamma_\mu e_R X^\mu$, 
$\overline{\ell}_L \gamma_\mu \ell_L X^\mu$ & $+/-$ & $180 (220)$ & $U(1)'$
\\
\hline
$(1,2,-\frac{3}{2})$ & $-1,-2$ & 
$\overline{e}_R \gamma_\mu \ell^c_L X^\mu$ & $+$ & $750(900)$ & $SU(3)_L \otimes U(1)_{X}$
\\
\hline
$(1,3,0)$ & $1,0,-1$ & 
$\overline{\ell}_L \gamma_\mu \ell_L X^\mu$
 & $+$ & $160(190)$ &  $SU(2)_1 \otimes SU(2)_2$
\\
\hline
$(\overline{3},1,-\frac{2}{3})$ & $-\frac{2}{3}$ & 
$\overline{e}_R \gamma_\mu d_R X^\mu$, 
$\overline{\ell}_L \gamma_\mu q_L X^\mu$
 & $+/-$  & $2000(2400)$ & $SU(4)_{C} \otimes U(1)_R$
\\
\hline
$(\overline{3},1,-\frac{5}{3})$ & $-\frac{5}{3}$ & 
$\overline{e}_R \gamma_\mu u_R X^\mu$
 & $+$ & $520(620)$ & $SU(4)_{C} \otimes U(1)_{R'}$
\\
\hline
$(3,2,\frac{1}{6})$ & $\frac{2}{3},-\frac{1}{3}$ & 
$\overline{\ell}_L \gamma_\mu u_R^c X^\mu$
  & $-$ & $\slash$ & $SU(5) \otimes U(1)_Z$ 
\\
\hline
$(3,2,-\frac{5}{6})$ & $-\frac{1}{3},-\frac{4}{3}$ & 
$\overline{e}_R \gamma_\mu q_L^c X^\mu$, 
$\overline{\ell}_L \gamma_\mu d_R^c X^\mu$
 & $+/-$ & $4400(5300)$ & $SU(5)$ 
\\
\hline
$(\overline{3},3,-\frac{2}{3})$ & $\frac{1}{3},-\frac{2}{3},-\frac{5}{3}$ & 
$\overline{\ell}_L \gamma_\mu q_L X^\mu$
 & $+$ & $540(650)$ & $SO(9) \otimes U(1)_R$
\\
\hline
  \end{tabular}
  \caption{\label{classificationVgm2v2} 
  List of new Lorentz vectors coupling to SM fermions at the renormalizable level and contributing to the $g-2$. 
  In the second column we provide the EM components of the $SU(2)_L$ multiplets, while 
  $\mathcal{O}^{g-2}_X$ denotes the $d=4$ operator responsible for the $g-2$ (gauge and flavor indices are 
  suppressed). 
  Representations with $Y=0$ are understood to be real. 
  For those cases where the contribution to the $(g-2)_\mu$ is non-negative we estimate in the fifth 
  column the 
  mass scale of the vector boson required in order to fit the discrepancy $\Delta a_\mu 
  = \left( 290 \pm 90 \right) \times 10^{-11}$ 
  (the number in the bracket corresponds to the $+1\sigma$ value). 
  For the estimate we take 
  the gauge coupling $g_X=1$ and an universal mass $M_X$ for all the components of the $SU(2)_L$ multiplets. 
  The last column displays the minimal embedding of the massive vector into an extended gauge group. 
  }
\end{table}

\subsection{Unitary gauge calculation}

Let us consider the Lagrangian in \eq{Lintgm2} with $k_Q=1$. 
The contribution to the muon anomalous magnetic moment (cf.~the two diagrams displayed in \fig{diagm2}) 
in the unitary gauge is known since long \cite{Leveille:1977rc} (see also Refs.~\cite{Queiroz:2014zfa,Jegerlehner:2009ry}). 
At the leading order in $m_\mu/M_X$ and $m_f/M_X$ it reads
\begin{multline}
\label{MFrena}
\Delta a_\mu^{(a)} = \frac{N_c Q_X}{4 \pi^2} \frac{m_\mu^2}{M_X^2} \left[ 
\abs{g_V}^2 \left(-\frac{5}{6} + \frac{m_f}{m_\mu} \right) 
+ \abs{g_A}^2 \left(-\frac{5}{6} - \frac{m_f}{m_\mu} \right) \right]  \\ 
= \frac{N_c Q_X}{4 \pi^2} \frac{m_\mu^2}{M_X^2} \left[ 
-\frac{5}{12} \left( \abs{g_L}^2 + \abs{g_R}^2 \right) 
+ \Re \left( g_L g^*_R\right) \frac{m_f}{m_\mu} \right] \, , 
\end{multline} 
\begin{multline}
\label{MFrenb}
\Delta a_\mu^{(b)} = \frac{N_c Q_f}{4 \pi^2} \frac{m_\mu^2}{M_X^2} \left[ 
\abs{g_V}^2 \left(\frac{2}{3} - \frac{m_f}{m_\mu} \right) 
+ \abs{g_A}^2 \left(\frac{2}{3} + \frac{m_f}{m_\mu} \right) \right]  \\ 
= \frac{N_c Q_f}{4 \pi^2} \frac{m_\mu^2}{M_X^2} \left[ 
\frac{1}{3} \left( \abs{g_L}^2 + \abs{g_R}^2 \right) 
- \Re \left( g_L g^*_R\right) \frac{m_f}{m_\mu} \right] \, ,
\end{multline} 
where $Q_{X,f}$ denote the EM charges of $X$ and $f$, while $N_c=3$ $(1)$ for color triplets (singlets).  
Note that in the second step of \eqs{MFrena}{MFrenb} we switched to the chiral basis couplings $g_L = g_V - g_A$ and 
$g_R = g_V + g_A$, which is a better language for $SU(2)_L \otimes U(1)_Y$ invariant interactions.
The generalization in flavor space for a generic gauge theory is also straightforward. The interaction term 
involving $X$, a muon and a SM fermion mass eigenstate field $f_i$ reads 
\beq 
\label{unitarystr}
\overline{\mu} \left( g_L U_L^{\mu i} \gamma_\alpha P_L + g_R U_R^{\mu i} \gamma_\alpha P_R \right) f_i X^\alpha + \text{h.c.} \, ,
\eeq
where $P_{L,R} = \tfrac{1}{2} (1 \mp \gamma_5)$ are chiral projectors and $U_{L,R}$ 
are unitary matrices in flavor space which perform the rotation from the flavor to the mass 
basis. Consequently, \eqs{MFrena}{MFrenb} are generalized into 
\begin{align}
\label{MFrenaGEN}
\Delta a_\mu^{(a)} &= \frac{N_c Q_X}{4 \pi^2} \frac{m_\mu^2}{M_X^2} \left[ 
-\frac{5}{12} \left( \abs{g_L}^2 + \abs{g_R}^2 \right) 
+ \Re \left( g_L g^*_R\right) \Re \left( U_L^{\mu i} U^{*\mu i}_R \right) \frac{m_{f_i}}{m_\mu} \right] \, , \\
\label{MFrenbGEN}
\Delta a_\mu^{(b)} &= \frac{N_c Q_f}{4 \pi^2} \frac{m_\mu^2}{M_X^2} \left[ 
\frac{1}{3} \left( \abs{g_L}^2 + \abs{g_R}^2 \right) 
- \Re \left( g_L g^*_R\right) \Re \left( U_L^{\mu i} U^{*\mu i}_R \right) \frac{m_{f_i}}{m_\mu} \right] \, ,
\end{align} 
where in the first term of the square brackets we exploited the 
unitarity relation $(U_{L,R} U^\dag_{L,R})^{\mu\mu}= 1$. On the other hand, the 
LR contribution in \eqs{MFrenaGEN}{MFrenbGEN} is weighted by the fermion mass $m_{f_i}$ and, 
depending on the specific UV gauge completion, 
by a generally unknown unitary matrix element.  
 
In reproducing the unitary gauge calculation we would like to mention a subtlety that one encounters 
when employing the unitary gauge at the loop level.\footnote{Another option could be that of using a different gauge, 
like the 't Hooft-Feynman gauge. 
} 
It is known that one should not shift momenta in more-than-logarithmically divergent integrals, 
otherwise spurious surface terms can change the final result by a finite amount 
(see e.g.~Chapter 6.2 in \cite{Cheng:1985bj}). This is a potential issue in the unitary gauge, since the degree of 
superficial divergence of the loop integrals gets worsened. Though the contribution to the $g-2$ 
must be finite in a renormalizable theory, one still needs to regularize the integrals in order not to 
meet the aforementioned issue. Indeed, we verified that the result of the calculation differs by a finite amount 
if one naively computes the integrals in $d=4$ dimensions, instead of using dimensional regularization in $d = 4 - 2 \epsilon$ 
and taking the $\epsilon \to 0$ limit at the very end.

\subsection{New vectors' contributions, gauge embeddings and bounds}
\label{vectorcases}

We proceed now by detailing the contribution of the new vectors in \Table{classificationVgm2v2} to the $(g-2)_\mu$ 
by using \eqs{MFrenaGEN}{MFrenbGEN} and estimate in turn the value of $M_X$ which is required in order to 
explain the discrepancy.  
Next, we discuss for each case a minimal gauge embedding of the new Lorentz vector. 
In order for the model to be phenomenologically viable, 
the SM fermions and Higgs boson must be properly embedded into the extended matter multiplets 
and the absence of gauge anomalies must be fulfilled. 
Regarding these last two points, we will not enter too much into details, but just refer to the existing literature 
when possible. For those cases where the SM matter embedding has not been discussed yet we will 
see that there exist independent arguments which are actually sufficient in order to exclude those 
possibilities as an explanation of the $(g-2)_\mu$. 
In particular, for any minimal viable realization we estimate indirect bounds from B and L number and 
flavor violating processes as well as limits from direct searches. 
To simplify the notation, we esplicitate the flavor structure only when needed. It is 
otherwise understood a unitary structure like that in \eq{unitarystr}, as the most general 
gauge interaction of the massive vector with the SM matter fields.

\subsubsection{$(1,1,0)$}
\label{X110}

Sticking to a flavor diagonal $Z'$, the interaction Lagrangian is 
\beq
\label{LX110}
\mathcal{L}^{g-2}_{\rm int} \supset
g_{X_1} \overline{e}_R\gamma_{\mu}e_{R}X^{\mu}
+g_{X_2}\overline{\ell}_{L}\gamma_{\mu} \ell_{L} X^{\mu} 
\supset g_{X_1} \overline{e}_R\gamma_{\mu}e_{R}X^{\mu}
+g_{X_2} \overline{e}_{L}\gamma_{\mu} e_{L}X^{\mu} \, , 
\eeq
which yields 
\beq 
\Delta a_{\mu} = - \frac{1}{12 \pi^2} \frac{m^2_\mu}{M^2_X} \left( g_{X_1}^2 + g_{X_2}^2 - 3 g_{X_1} g_{X_2} \right) \, .
\eeq
The latter is positive for 
$\frac{1}{2}(3-\sqrt{5}) < g_{X_2}/g_{X_1} < \frac{1}{2}(3+\sqrt{5})$, while it reaches its maximal positive value 
\beq 
\label{DamuX110max}
\Delta a_{\mu}^{\rm max} = \frac{g_{X_1}^2}{12 \pi^2} \frac{m^2_\mu}{M^2_X} \frac{5}{4} \, ,
\eeq
for $g_{X_2}/g_{X_1} = 3/2$.\footnote{This option is prone to gauge anomaly cancellation 
constraints, since the couplings are chiral. 
However, anomalies can be fixed by coupling $X$ to another fermionic sector. 
Alternatively, one can consider the anomaly free scenario $g_{X_1} = g_{X_2}$. 
In such a case, the required vector boson mass is $M_X / g_{X_1} = 180$ GeV.} 
From \eq{DamuX110max} we find that the $(g-2)_\mu$ requires 
$M_X / g_{X_1} = 200$ GeV. 

The gauge embedding corresponds to that of an extra $U(1)'$ factor and the lower 
bounds on $M_X$ are quite model dependent. For instance, in the case of a sequential SM $Z'$, 
ATLAS \cite{ATLAS:2015} and CMS \cite{CMS:2015nhc} set the bound respectively to  $M_{Z'} > 3.4$ TeV and $M_{Z'} > 3.2$ TeV by looking into di-lepton channels. On the other hand, even if the $Z'$ couples only to muons 
(as minimally required by the muon $g-2$), neutrino trident production $\nu_\mu N \to \nu_\mu N \mu^+ \mu^-$ 
from CCFR data \cite{Mishra:1991bv} 
rules out the explanation of the $(g-2)_\mu$ anomaly for masses $M_{Z'} \gtrsim 400$ MeV \cite{Altmannshofer:2014pba}, 
while the available low-mass range can be covered at future neutrino beam facilities. 

Without requiring additional exotic fermions contributing to the $(g-2)_\mu$, there are two other options leading to a 
viable $Z'$ explanation of the $(g-2)_\mu$. The first one is a dark photon or $Z$ without direct couplings to the 
SM fields, which can still contribute to the $(g-2)_\mu$ via a gauge kinetic mixing to the EM 
current \cite{Gninenko:2001hx,Pospelov:2008zw,Davoudiasl:2014kua}. 
As shown in Ref.~\cite{Davoudiasl:2014kua}, the explanation in terms of a light gauge boson of $\mathcal{O}(100)$ 
MeV requires however a sizeable invisible decay channel of the $Z'$. 
Another possibility, recently discussed in Refs.~\cite{Heeck:2016xkh,Altmannshofer:2016brv}, 
is that of a flavor off-diagonal coupling of the $Z'$ to the $\mu$ and $\tau$ sector. This can explain the 
$(g-2)_\mu$ for a $Z'$ heavier than the $\tau$ lepton, compatibly with all the existing bounds.

\subsubsection{$(1,2,-\frac{3}{2})$}
\label{X12m32}

Let us consider the interaction Lagrangian
\begin{align}
\mathcal{L}^{g-2}_{\rm int} &\supset g_X\overline{e}_{R}\gamma_{\mu}\ell_{L}^c X^{\mu}+\text{h.c.}
=g_X \left[\overline{e}_{R}\gamma_{\mu}\nu_{L}^c X_{-1}^{\mu}
+ \overline{e}_{R}\gamma_{\mu}e_{L}^c X_{-2}^{\mu}\right] +\text{h.c.} \nonumber \\
& = 
g_X \left[ \overline{e}_{R}\gamma_{\mu}\nu_{L}^c X_{-1}^{\mu}
+ \left( 
\tfrac{1}{2} \underbrace{\overline{e} \gamma_{\mu} C \overline{e}^T}_{=0} X_{-2}^{\mu} 
+\tfrac{1}{2} \overline{e} \gamma_{\mu} \gamma_5 C \overline{e}^T X_{-2}^{\mu} \right) \right]
+\text{h.c.}
\, ,
\end{align}
where in the last step we have emphasized the fact that the vector current associated to the doubly-charged component of $X$ is zero 
by symmetry reasons.\footnote{In fact, by using the anticommuting properties of fermion fields, 
$C \gamma^{\mu T} C^{-1} = - \gamma^\mu$ and $C^T = - C$ 
one gets: 
$\overline{e} \gamma_{\mu} C \overline{e}^T 
= (\overline{e} \gamma_{\mu} C \overline{e}^T)^T
= - \overline{e} C^T \gamma_{\mu}^T \overline{e}^T 
= - \overline{e} \gamma_{\mu} C \overline{e}^T
$.
 } 
Note, also, that the Feynman rule of $X_{-2}$ features an extra 2 symmetry factor 
in the $\mu \mu X$ vertex (and hence a factor 4 in the $g-2$ amplitude). At the end 
the final contribution to the $(g-2)_\mu$ is found to be 
\beq 
\Delta a_{\mu}= \frac{23}{16\pi^2}\frac{m_{\mu}^{2}}{M_{X}^2}g_X^2  \, , 
\eeq
and in order to reproduce the $(g-2)_\mu$ we need $M_X / g_X = 750$ GeV.  

The gauge embedding in this case is minimally realized via the so-called 331 models, which are 
based on the extended gauge group $SU(3)_C \otimes SU(3)_L \otimes U(1)_X$ \cite{Pisano:1991ee}.  
The SM hypercharge is embedded via the relation 
\beq 
Y = \xi T^8_L + X \, ,  
\eeq
where $T^8_L$ is a Cartan generator of the $SU(3)_L$ algebra (normalized as $\Tr T^{a}_L T^{b}_L = \tfrac{1}{2} \delta^{ab}$) 
and the parameter $\xi$ defines a class of different models (see e.g.~\cite{Okada:2016whh}), 
while the $X$-charge assignment of the matter fields defines the embedding 
of the SM fermions into the extended matter multiplets.  
In particular, in order to obtain $(1,2,-\tfrac{3}{2})$ as a would-be goldstone boson (WBG) 
one needs $\xi = \pm \sqrt{3}$. On top of that, the $SU(3)_C \otimes SU(3)_L \otimes U(1)_X \to SU(3)_C \otimes SU(2)_L \otimes U(1)_Y$ 
breaking also delivers a $Z'$. 

Ref.~\cite{Kelso:2014qka} studied the interplay between the $(g-2)_\mu$ and the electroweak and collider 
constraints in different classes of 331 models and found that no renormalizable extension can explain the 
$(g-2)_\mu$, mainly due to lower bounds on the $Z'$ mass which translate into lower bounds on the 
singly and doubly charged components of $(1,2,-\tfrac{3}{2})$ within the specific models. 

\subsubsection{$(1,3,0)$}
\label{X130}

In this case we use a matrix representation for the (real) electroweak triplet
\beq 
\mathbf{X}^\mu = \frac{\sigma^i X^{i\mu}}{\sqrt{2}} = 
\left( 
\begin{array}{cc}
\frac{X_0^\mu}{\sqrt{2}} & X_{+1}^\mu \\
X_{-1}^\mu & - \frac{X_0^\mu}{\sqrt{2}} 
\end{array}
\right) \, ,
\eeq
and the relevant Lagrangian for the $(g-2)_\mu$ is 
\beq
\mathcal{L}^{g-2}_{\rm int} \supset g_X \overline{\ell}_{L}\gamma_{\mu} \mathbf{X}^\mu \ell_{L} 
\supset - \frac{g_X}{\sqrt{2}} \overline{e}_{L} \gamma_{\mu} e_{L} X_0^\mu 
+ g_X \left( \overline{e}_{L} \gamma_{\mu} \nu_{L} X_{-1}^\mu +\text{h.c.} \right) \, .
\eeq
The contribution to $\Delta a_{\mu}$ is 
\beq 
\Delta a_{\mu}= \frac{1}{16\pi^2}\frac{m_{\mu}^{2}}{M_{X}^2}g_X^2 \, ,
\eeq
from which we get that in order to reproduce the $(g-2)_\mu$ we need $M_X / g_X = 160$ GeV.

A minimal gauge extensions delivering $(1,3,0)$ as a WBG is given by $SU(2)_1 \otimes SU(2)_2$, 
spontaneously broken to the diagonal subgroup, which is identified with $SU(2)_L$. 
Different variant models depend on the SM fermions' embedding. 
Let us mention, for instance, the ``un-unified" model 
where left-handed quarks and leptons are respectively assigned to $SU(2)_1$ and $SU(2)_2$ 
\cite{Georgi:1989ic,Georgi:1989xz}, 
and the ``non-universal" model in which the third generation left-handed fermions undergo a different $SU(2)$ 
interaction from those of the first two generations \cite{Malkawi:1996fs}. 
Due to the symmetry breaking pattern the masses of the $W'$ and $Z'$ contained in the $(1,3,0)$ 
are quite degenerate and their mixing with the $W$ and $Z$ leads to 
strong constraints from precision electroweak measurements.
In fact, a global analysis including $Z$-pole observables, $W$ properties, $\tau$ lifetime, $\nu N(e)$-scattering and 
atomic parity violation sets the bound 
at the level of $M_{W'} \sim M_{Z'} \gtrsim 2.5$ TeV \cite{Hsieh:2010zr}. 
On the other hand, the new charged vector bosons can be pair-produced and leave a signature of leptons and missing energy.
By recasting LHC slepton searches \cite{Khachatryan:2014qwa}, Ref.~\cite{Freitas:2014pua} 
sets the lower bound $M_{W'} \gtrsim 400$ GeV, which holds 
irrespectively of the UV completion. 
This clearly rules out the possible explanation of the $(g-2)_\mu$ discrepancy.

\subsubsection{$(\overline{3},1,-\frac{2}{3})$}
\label{X31m23}

The interaction Lagrangian can be written as 
\beq
\label{LagX31m23}
\mathcal{L}^{g-2}_{\rm int} \supset 
g_X \left( \overline{e}_R\gamma_{\mu}d_{R}X^{\mu}
+ \overline{\ell}_{L}\gamma_{\mu} q_{L}X^{\mu} \right) +\text{h.c.}
\supset 
g_X \left( 
\overline{e}_{R\mu} U_R^{\mu i} \gamma_{\mu} d_{Ri} X^{\mu}
+ \overline{e}_{L\mu} U_L^{\mu i} \gamma_{\mu} d_{Li}X^{\mu} \right)  +\text{h.c.} \, ,
\eeq
where $U_{L,R}$ are unitary mixing matrices. 
The contribution to the $(g-2)_\mu$ is then
\beq
\Delta a_{\mu}= 
\frac{1}{4\pi^2}\frac{m_{\mu}^{2}}{M_{X}^2} g_X^2 \left( 1 - \Re \left( U_L^{\mu i} U^{*\mu i}_R \right) \frac{m_{d_i}}{m_\mu} \right) \, .
\eeq
In order to maximize the contribution, we assume maximal mixing in the bottom direction with $\Re \left( U_L^{\mu i} U^{*\mu i}_R \right)=-1$, 
thus inferring $M_X / g_X = 2.0$ TeV in order to explain the $(g-2)_\mu$ discrepancy. 

The minimal UV completion of the $(\overline{3},1,-\frac{2}{3})$ vector leptoquark is given by 
the quark-lepton unification model based on the gauge group 
$SU(4)_C \otimes SU(2)_L \otimes U(1)_R$ (see e.g.~\cite{Perez:2013osa}), 
which is a particular case of the more general Pati-Salam group \cite{Pati:1974yy}. 
The SM hypercharge is embedded via the relation
\beq
Y = \frac{\sqrt{6}}{3} T^{15}_C + R \, , 
\eeq
where $T^{15}_C$ is a properly normalized Cartan generator of $SU(4)_C$ algebra ($\Tr T^{a}_C T^{b}_C = \tfrac{1}{2} \delta^{ab}$). 
The $R$-charge assignment of the matter fields defines the embedding 
of the SM fermions into the extended matter multiplets. 
On top of $(\overline{3},1,-\frac{2}{3})$, the $SU(4)_C \otimes SU(2)_L \otimes U(1)_R \to SU(3)_C \otimes SU(2)_L \otimes U(1)_Y$ 
breaking also delivers a $Z'$ as a WBG. 

The vector leptoquark $(\overline{3},1,-\frac{2}{3})$ contributes to the rare decay $K^0_L \to e^{\mp} \mu^{\mp}$, 
which for $\mathcal{O}(1)$ couplings yields the bound 
$M_X \gtrsim 10^3$ TeV (see e.g.~\cite{Valencia:1994cj,Smirnov:2007hv}). 
Such a strong constraint can be in principle evaded if one takes into account the freedom 
in the flavor mixing between quarks and leptons, 
due to the unknown unitarity matrices $U_{L,R}$ in \eq{LagX31m23}. 
In such a case, a full set of observables from rare $K$ and $B$ meson decays must be taken into account and, by combining the strongest constraints, Refs.~\cite{Kuznetsov:2012ai,Kuznetsov:2012ad} find 
$M_X \gtrsim 38$ TeV, regardless of flavor mixing. 
Remarkably, a numerical scan of the multi-dimensional 
parameter space reveals the existence of viable configurations with masses as low as 
$M_X \sim 12$ TeV \cite{BBDL2016}, which is however still too high for the explanation of the $(g-2)_\mu$. 

\subsubsection{$(\overline{3},1,-\frac{5}{3})$}
\label{X31m53}

From the interaction Lagrangian 
\begin{equation}
\mathcal{L}^{g-2}_{\rm int} \supset g_X\overline{e}_{R}\gamma_{\mu}u_{R}X^{\mu} +\text{h.c.} \, ,
\end{equation}
the contribution to $\Delta a_{\mu}$  is found to be 
\beq 
\Delta a_{\mu}=\frac{11}{16\pi^2}\frac{m_{\mu}^2}{M_{X}^2}g_X^2 \, .
\eeq
The mass scale required to fit the $(g-2)_\mu$ is $M_X / g_X = 520$ GeV. 

The gauge extension of this case is analogous to the previous one, and is given by 
the $SU(4)_C \otimes SU(2)_L \otimes U(1)_{R'}$ group, whose breaking 
also delivers an extra $Z'$ as a WBG.
The corresponding embedding of the SM hypercharge is
\beq
Y = \frac{5}{2}\frac{\sqrt{6}}{3} T^{15}_C + R' \, .
\eeq
The $R'$-charges of the matter fields define the embedding 
of the SM fermions into the extended matter multiplets. 
The latter differs substantially from the standard Pati-Salam embedding 
and we did not attempt to build a realistic fermionic sector. 
However, even without discussing that, such a light $M_X$ (as required by the $(g-2)_\mu$) 
is ruled out by collider searches. 

In order to show that let us make explicit the unitary structure of the leptoquark interactions in 
flavor space
\beq 
g_X U_R^{ij} \overline{e}_{iR}\gamma_{\mu}u_{jR}X^{\mu} \, ,
\eeq
where the (a priori unknown) unitary matrix $U_R$ controls the branching ratios of $X \to e_i u_j$. 
In particular, we have 
\begin{align} 
\label{brXtoej}
\mathcal{B}(X \to e j) &= \frac{\sum_{j=u,c} \abs{U_R^{ej}}^2}{\sum_{i=e,\mu,\tau} \sum_{j=u,c,t} \abs{U_R^{ij}}^2} 
= \frac{1-\abs{U_R^{et}}^2}{3} \, , \\
\label{brXtomuj}
\mathcal{B}(X \to \mu j) &= \frac{\sum_{j=u,c} \abs{U_R^{\mu j}}^2}{\sum_{i=e,\mu,\tau} \sum_{j=u,c,t} \abs{U_R^{ij}}^2} 
= \frac{1-\abs{U_R^{\mu t}}^2}{3} \, .
\end{align}
On the other hand, the pair-production cross section of $X$ is unambiguously fixed by QCD and we can use the 
CMS searches in Ref.~\cite{Khachatryan:2015vaa} in order to constrain the combined $X \to e j$ and $X \to \mu j$ 
channels. Note that the elements $U_R^{ej}$ and $U_R^{\mu j}$ are still related by unitarity, 
and even in the worse case scenario where the top is maximally mixed with the first two generation leptons 
(thus leading to a potential reduction of the branching ratios in \eqs{brXtoej}{brXtomuj}), we can parametrize 
the mixing matrix elements as $U_R^{ej} =\sin\phi$ and $U_R^{\mu j} =\cos\phi$. 
The most conservative bound is obtained by simultaneously minimizing the two branching ratios, since 
$ej$ and $\mu j$ searches lead to similar bounds.  
This is obtained by taking $\phi = \pi / 4$, which corresponds to a $\mathcal{B}$ of $1/6$ in both the channels. 
By simply rescaling the cross sections in Figs.~13 and 14 of Ref.~\cite{Khachatryan:2015vaa} by a $(1/6)^2$ factor 
we obtain $M_X \gtrsim 1$ TeV, which is sufficient in order to exclude the explanation of the $(g-2)_\mu$ in terms of 
$(\overline{3},1,-\frac{5}{3})$.

\subsubsection{$(3,2,\frac{1}{6})$}
\label{X3216}

Given the interaction Lagrangian
\beq
\mathcal{L}^{g-2}_{\rm int} \supset g_X \overline{\ell}_L \gamma_{\mu}u^{c}_{R} X^{\mu} +\text{h.c.}
\supset g_X \overline{e}_L \gamma_{\mu} u^{c}_{R} X_{-1/3}^{\mu} +\text{h.c.}
\, ,
\eeq
the contribution to $\Delta a_{\mu}$ is 
\beq
\Delta a_{\mu}=-\frac{1}{16\pi^2}\frac{m_{\mu}^2}{M_X^2}g_X^2 \, ,
\eeq
which, being negative, cannot explain the $(g-2)_\mu$. 

For completeness, we mention that this case corresponds to the ``flipped" SU(5) embedding 
of the SM hypercharge \cite{DeRujula:1980qc,Barr:1981qv}. 
Moreover, the breaking also delivers an extra $Z'$ as a WBG.

\subsubsection{$(3,2,-\frac{5}{6})$}
\label{X32m56}

The interaction Lagrangian can be written as 
\begin{align}
\mathcal{L}^{g-2}_{\rm int}& \supset g_X 
\left( \overline{e}_R \gamma_\mu q_L^c X^\mu 
+ \overline{\ell}_L \gamma_\mu d_R^c X^\mu \right) +\text{h.c.} \nonumber \\
& \supset 
g_X \left( \overline{e}_{R\mu} \tilde{U}_R^{\mu i} \gamma_\mu u_{Li}^c X^\mu_{-1/3} 
+ \overline{e}_{R\mu} U_R^{\mu i} \gamma_\mu d_{Li}^c X^\mu_{-4/3} 
+ \overline{e}_{L\mu} U_L^{\mu i}  \gamma_\mu d_{Ri}^c X^\mu_{-4/3} \right)
+\text{h.c.} \, ,
\end{align}
where $\tilde{U}_R$ and $U_{L,R}$ are unitary mixing matrices. 
The contribution to $\Delta a_{\mu}$ is found to be 
\beq
\Delta a_{\mu}= 
\frac{5}{16\pi^2}\frac{m_{\mu}^2}{M_{X}^2} g_X^2
\left( 3 - 4 \, \Re \left( U_L^{\mu i} U^{*\mu i}_R \right) \frac{m_{d_i}}{m_\mu} \right) \, .
\eeq
In order to maximize the contribution, we assume maximal mixing in the bottom direction with 
$\Re \left( U_L^{\mu i} U^{*\mu i}_R \right)=-1$, and thus we get $M_X / g_X = 4.4$ TeV 
in order to explain the $(g-2)_\mu$ discrepancy. 

The UV completion of this vector leptoquark is the standard $SU(5)$ \cite{Georgi:1974sy}, which 
clearly rules out the interpretation of the $(g-2)_\mu$, since $M_X \gtrsim 10^{15}$ GeV from proton decay and unification constraints.

\subsubsection{$(\overline{3},3,-\frac{2}{3})$}
\label{X33m23}

By using the following electroweak triplet matrix representation
\beq 
\mathbf{X}^\mu = \frac{\sigma^i X^{i\mu}}{\sqrt{2}} = 
\left( 
\begin{array}{cc}
\frac{X_{-2/3}^\mu}{\sqrt{2}} & X_{+1/3}^\mu \\
X_{-5/3}^\mu & - \frac{X_{-2/3}^\mu}{\sqrt{2}} 
\end{array}
\right) \, ,
\eeq
we can write the interaction Lagrangian as 
\beq
\mathcal{L}^{g-2}_{\rm int} \supset g_X \overline{\ell}_{L}\gamma_{\mu} \mathbf{X}^\mu q_{L} +\text{h.c.} 
\supset 
- \frac{g_X}{\sqrt{2}} \overline{e}_{L} \gamma_{\mu} d_{L} X_{-2/3}^\mu 
+ g_X \overline{e}_{L} \gamma_{\mu} u_{L} X_{-5/3}^\mu +\text{h.c.} \, .
\eeq
This leads to 
\beq 
\Delta a_{\mu}= \frac{3}{4\pi^2}\frac{m_{\mu}^2}{M_X^2} g_X^2 \, ,
\eeq
which implies $M_X / g_X = 540$ GeV for the explanation of the $(g-2)_\mu$ discrepancy. 

On top of possible collider searches which we do not discuss, the main no-go here is the gauge embedding which 
requires the $SU(3)_C$ and $SU(2)_L$ SM gauge factor to get unified below the TeV scale, which is clearly 
ruled out. 

For completeness, we provide a symmetry breaking pattern delivering $(\overline{3},3,-\frac{2}{3})$ 
as a WBG. The minimal option we were able to find is 
$SO(9) \otimes U(1)_R \to SU(4)_C \otimes SU(2)_L \otimes U(1)_R \to SU(3)_C \otimes SU(2)_L \otimes U(1)_Y$. 
Here, the branching rule of the adjoint under $SO(9) \to SU(4)_C \otimes SU(2)_L$ 
is given by $36 \to (1,3) \oplus (6,3) \oplus (15,1)$ \cite{Feger:2012bs}. 
Next, under $SU(4)_C \otimes SU(2)_L \otimes U(1)_R \to SU(3)_C \otimes SU(2)_L \otimes U(1)_Y$, 
$(6,3,0) \to (3,3,\tfrac{2}{3}) \oplus (\overline{3},3,-\frac{2}{3})$, provided the embedding of the SM hypercharge 
is $Y = \frac{2\sqrt{6}}{3} T^{15}_C + R$.  
On the other hand, the embedding of the SM fermions is non-trivial and we did not attempt to build a realistic model.

\section{Discussion and conclusions}
\label{concl}

The increase of the degree of divergence of loop diagrams in presence of non-gauge massive vectors is something well-known.
A typical example is given by meson mixing amplitudes for which the box diagrams involving massive vectors, 
with propagators as in \eq{propcan}, are quadratically divergent (see e.g.~\cite{Davidson:1993qk,Dorsner:2016wpm})
\beq 
\Delta m^{ij}_M \propto \Lambda^2 \sum_{f,f'} U^{if} U^{*jf} U^{if'} U^{*jf'} \, . 
\eeq
Here, $M$ denotes a $K$, $D$ or $B$ meson, $ij$ are the meson constituent quarks and $ff'$ the fermions exchanged in the loop. 
In a gauge theory the $U$-matrices are unitary and a GIM-like mechanism ensures the cancellation 
of the quadratic divergence, as it should in a renormalizable theory. Yet another example is given by the divergent contributions 
to electroweak precision observables from composite vectors (see 
e.g.~\cite{Barbieri:2008cc,Cata:2010bv,Orgogozo:2011kq,Pich:2012jv,Contino:2015mha}).
In a similar way, we have seen that the triple vector boson vertex in diagram $(a)$ of \fig{diagm2} is the 
origin of the logarithmic divergence of the $g-2$. 
This is also to be expected, since renormalizability crucially hinges on the exact values of the non-abelian vertices. 

In this paper we have classified all the possible quantum numbers of a new massive vector which can couple 
to SM fields via $d \leq 4$ operators (cf.~\Tables{classificationVSMfermions}{classificationVSMbosons}). 
Only a subset of these irreps can contribute to the $g-2$, and for each of them we provided the embedding 
of the massive vector into a spontaneously broken gauge theory (cf.~\Table{classificationVgm2v2}). 
While some gauge extensions are of course well-known, those concerning $(\overline{3},1,-\tfrac{5}{3})$ and 
$(\overline{3},3,-\tfrac{2}{3})$ are to our knowledge new. 
The maybe less obvious result of this paper is that after embedding the massive vector into an extended 
gauge symmetry, such that the $g-2$ can be unambiguously computed in terms of a renormalizable Lagrangian, renormalizability highly constrains the interactions 
of the vector field. In fact, a combination of direct and indirect bounds, as well as unification constraints, 
rules out the possible explanation of the muon $g-2$ in terms of new massive vectors, 
with the only notable exception of an abelian gauge extension. The latter, indeed, is less constrained 
because the extra gauge group is factorized with respect to the SM gauge group and the couplings to SM 
fields are highly model dependent. 

It should be also stressed that the starting hypothesis of a 1-particle 
vector extension of the SM is often violated in the renormalizable case, since new sectors of the theory are often required 
by the consistency of the symmetry breaking pattern (e.g.~scalar fields breaking the extended symmetry, extra WGBs 
and new fermions fitting the extended matter multiplets) and they cannot be arbitrary decoupled from the new vector 
mass scale. In principle, the inclusion of these extra fields can provide extra contribution for explaining the $(g-2)_\mu$. 
This, however, is model dependent and goes beyond the original question.

Finally, we would like to comment on a couple of other phenomenologically relevant contexts where similar 
observations apply as in the $g-2$ case. 
The first one is that of $B$-meson decay anomalies. New massive vectors have been recently proposed for addressing 
some $3$$\sigma$ level discrepancies in semileptonic $B$-meson decays \cite{Aaij:2013qta,Aaij:2014ora,Aaij:2015yra}. 
Aside from abelian gauge extensions (see e.g.~\cite{Descotes-Genon:2013wba,Gauld:2013qba,Altmannshofer:2014cfa,Sierra:2015fma,Belanger:2015nma,Crivellin:2015era}) there are three non-trivial irreps which are well-suited for addressing 
$B$-meson anomalies if they couple to SM fermions exclusively via left-handed currents: 
$(1,3,0)$ \cite{Greljo:2015mma,Calibbi:2015kma,Boucenna:2016wpr,Boucenna:2016qad}, 
$(\overline{3},1,-\frac{2}{3})$ \cite{Alonso:2015sja,Barbieri:2015yvd} and 
$(\overline{3},3,-\frac{2}{3})$ \cite{Fajfer:2015ycq,Buttazzo:2016kid}. 
In these examples the issue of renormalizability 
was not central, being all the main experimental anomalies to be explained at tree level 
(see however Ref.~\cite{Barbieri:2015yvd} for a discussion of divergent loop observables). 
Nevertheless, if one requires these non-abelian massive vectors to arise from a spontaneously broken extended 
gauge symmetry new extra constraints must be fulfilled. 
We already discussed a minimal gauge embedding for each of these three vector irreps  
in \sect{vectorcases}. As far as regards $(1,3,0)$, if it couples universally to the three SM families,  
the unitarity of the gauge interactions forces the neutral currents to be diagonal in flavor space and the charged currents to be aligned to the SM, 
thus lacking of the required amount of flavor violation for $b \to s$ and $b \to c$ transitions. 
As pointed out in Refs.~\cite{Boucenna:2016wpr,Boucenna:2016qad}, a viable UV gauge completion of $(1,3,0)$ 
for the explanation of the the $B$-anomalies requires universal gauge couplings 
and an extra source of flavor violation, e.g.~from the mixing of the SM quarks with new vector-like fermions. 
Similarly, in the case of $(\overline{3},1,-\frac{2}{3})$ the unitary structure of the 
leptoquark interactions with the SM fermions is such that a bunch of rare processes from rare $K$ and $B$ meson decays 
cannot be simply set to zero by switching-off right-handed currents. As discussed in \sect{X31m23}, the mass of the new 
vector is bounded to lie in the multi-tens of TeV region and hence too high in order to explain all the $B$ anomalies. 
Finally, the case of a light $(\overline{3},3,-\frac{2}{3})$ is also trivially excluded, since if it were to come from a gauge theory 
the strong and electroweak couplings would have to be unified at the TeV scale. 

Massive vectors mediators have been also recently invoked for the explanation of the LHC 
di-photon excess 
(see e.g.~\cite{Murphy:2015kag,deBlas:2015hlv}). 
In such a case both the production of the scalar resonance via gluon
fusion and its decay into two photons 
is obtained via a loop of massive vectors featuring triple and quartic vector boson vertices, 
which lead in general to divergent contributions. 
On the other hand, by sticking to a finite result for the loop functions in order to fit the cross-section signal 
one is implicitly assuming that the vector boson has a gauge origin and, as we saw in the previous examples, 
it is non-trivial to satisfy all the relevant bounds in presence of a gauge vector mediator at the TeV scale. 

\section*{Acknowledgments}

We thank Michele Frigerio, Jernej F.~Kamenik and Marco Nardecchia for helpful discussions. 
The work of C.B.~and L.D.L.~is supported by the Marie Curie CIG program, project number PCIG13-GA-2013-618439. 
The research of M.B.~is supported in part by the Swiss National Science Foundation (SNF) under contract 200021-159720.

\appendix

\section{1-particle vector extensions of the SM}
\label{classX}

In this Appendix we provide the classification of all the possible gauge quantum numbers of a Lorentz vector, $X^\mu$, 
which can couple to SM fields at the renormalizable level. We start by collecting in \Table{classificationVSMfermions} 
those cases where the new vector couples to SM fermions.  
In \Table{classificationVSMbosons} we classify instead $d \leq 4$ operators involving $X^\mu$ and SM bosons (either scalar or vector).

\begin{table}[htbp]
\renewcommand{\arraystretch}{1.4}
\centering
\begin{tabular}{@{} |c|c|c| @{}}
\hline
$X^\mu$ &  $Q_{\rm EM}$ &  $\mathcal{O}^{d=4}_X$   
\\ 
\hline
\hline
$(1,1,0)$ & 0 & 
$\overline{e}_R \gamma_\mu e_R X^\mu$, 
$\overline{\ell}_L \gamma_\mu \ell_L X^\mu$, 
$\overline{u}_R \gamma_\mu u_R X^\mu$,
$\overline{d}_R \gamma_\mu d_R X^\mu$,
$\overline{q}_L \gamma_\mu q_L X^\mu$ \\
\hline
$(1,1,1)$ & $1$ & 
$\overline{u}_R \gamma_\mu d_R X^\mu$ \\
\hline
$(1,2,-\frac{3}{2})$ & $-1,-2$ & 
$\overline{\ell}_L \gamma_\mu (e_R)^c X^\mu$ \\
\hline
$(1,3,0)$ & $1,0,-1$ & 
$\overline{\ell}_L \gamma_\mu \ell_L X^\mu$, 
$\overline{q}_L \gamma_\mu q_L X^\mu$ \\
\hline
$(\overline{3},1,-\frac{2}{3})$ & $-\frac{2}{3}$ & 
$\overline{e}_R \gamma_\mu d_R X^\mu$, 
$\overline{\ell}_L \gamma_\mu q_L X^\mu$ \\
\hline
$(\overline{3},1,-\frac{5}{3})$ & $-\frac{5}{3}$ & 
$\overline{e}_R \gamma_\mu u_R X^\mu$ \\
\hline
$(3,2,\frac{1}{6})$ & $\frac{2}{3},-\frac{1}{3}$ & 
$\overline{\ell}_L \gamma_\mu (u_R)^c X^\mu$, 
$\overline{q}^c_L \gamma_\mu d_R X^\mu$ \\
\hline
$(3,2,-\frac{5}{6})$ & $-\frac{1}{3},-\frac{4}{3}$ & 
$\overline{e}_R \gamma_\mu q^c_L X^\mu$, 
$\overline{\ell}_L \gamma_\mu d^c_R X^\mu$, 
$\overline{q}^c_L \gamma_\mu u_R X^\mu$ \\
\hline
$(\overline{3},3,-\frac{2}{3})$ & $\frac{1}{3},-\frac{2}{3},-\frac{5}{3}$ & 
$\overline{\ell}_L \gamma_\mu q_L X^\mu$ \\
\hline
$(\overline{6},2,\frac{1}{6})$ & $\frac{2}{3},-\frac{1}{3}$ &  
$\overline{q}^c_L \gamma_\mu d_R X^\mu$ \\
\hline
$(\overline{6},2,-\frac{5}{6})$ & $-\frac{1}{3},-\frac{4}{3}$ & 
$\overline{q}^c_L  \gamma_\mu u_R X^\mu$  \\
\hline
$(8,1,0)$ & 0 & 
$\overline{u}_R \gamma_\mu u_R X^\mu$,
$\overline{d}_R \gamma_\mu d_R X^\mu$,
$\overline{q}_L \gamma_\mu q_L X^\mu$ \\
\hline
$(8,1,1)$ & 1 & 
$\overline{u}_R \gamma_\mu d_R X^\mu$  \\
\hline
$(8,3,0)$ & $1,0,-1$ & 
$\overline{q}_L \gamma_\mu q_L X^\mu$ \\
\hline
  \end{tabular}
  \caption{\label{classificationVSMfermions} 
  List of new Lorentz vectors with $d=4$ coupling to SM fermions. 
  The EM charges of the particles in the multiplet and the relevant $d=4$ operators 
  are displayed (gauge and flavor indices are understood). 
    }
\end{table}

\begin{table}[htbp]
\renewcommand{\arraystretch}{1.4}
\centering
\begin{tabular}{@{} |c|c|c| @{}}
\hline
$\mathcal{O}_X$ &  $\text{dim}(\mathcal{O}_X)$ &  $X^\mu$ 
\\ 
\hline
\hline
$H D_\mu X^\mu$ 
& 3
& $(1,2,\frac{1}{2})$ \\
\hline
$H^\dag D_\mu X^\mu$ 
& 3
& $(1,2,-\frac{1}{2})$ \\
\hline
$HH D_\mu X^\mu$ 
& 4
& $(1,3,-1)$ \\
\hline
$HH^\dag D_\mu X^\mu$ 
& 4
& $(1,1 \oplus 3,0)$ \\
\hline
$HH X_\mu X^\mu$ 
& 4
& $(R,2k,-\frac{1}{2})$ \\
\hline
$HH^\dag X_\mu X^\mu$ 
& 4
& $(R,2k,0)$ \\
\hline
$HH^\dag X_\mu X^{\dag\mu}$ 
& 4
& $(C,n,Y)$ \\
\hline
$D_\mu X^\dag_{\nu} D^\mu X^{\nu}$ 
& 4
& $(C,n,Y)$ \\
\hline
$D_\mu X^\dag_{\nu} D^\nu X^{\mu}$ 
& 4
& $(C,n,Y)$ \\
\hline
$G_{\mu\nu} D^\mu X^{\nu}$ 
& 4
& $(8,1,0)$ \\
\hline
$W_{\mu\nu} D^\mu X^{\nu}$ 
& 4
& $(1,3,0)$ \\
\hline
$B_{\mu\nu} \partial^\mu X^{\nu}$ 
& 4
& $(1,1,0)$ \\
\hline
$G_{\mu\nu} X^\mu X^{\dag\nu}$ 
& 4
& $(C_{\neq 1},n,Y)$ \\
\hline
$W_{\mu\nu} X^\mu X^{\dag\nu}$ 
& 4
& $(C,n_{\neq 1},Y)$ \\
\hline
$B_{\mu\nu} X^\mu X^{\dag\nu}$ 
& 4
& $(C,n,Y)$ \\
\hline
$D_\mu X_\nu X^\mu X^{\nu}$ 
& 4
& $(C,2k+1,0)$ \\
\hline
$D_\mu X_\nu X^\mu X^{\dag\nu}$ 
& 4
& $(R,2k+1,0)$ \\
\hline
$\epsilon_{\mu\nu\rho\sigma} G^{\mu\nu} D^\rho X^{\sigma}$ 
& 4
& $(8,1,0)$ \\
\hline
$\epsilon_{\mu\nu\rho\sigma} W^{\mu\nu} D^\rho X^{\sigma}$ 
& 4
& $(1,3,0)$ \\
\hline
$\epsilon_{\mu\nu\rho\sigma} B^{\mu\nu} \partial^\rho X^{\sigma}$ 
& 4
& $(1,1,0)$ \\
\hline
$\epsilon_{\mu\nu\rho\sigma} G^{\mu\nu} X^\rho X^{\dag\sigma}$ 
& 4
& $(C_{\neq 1},n,Y)$ \\
\hline
$\epsilon_{\mu\nu\rho\sigma} W^{\mu\nu} X^\rho X^{\dag\sigma}$ 
& 4
& $(C,n_{\neq 1},Y)$ \\
\hline
$\epsilon_{\mu\nu\rho\sigma} B^{\mu\nu} X^\rho X^{\dag\sigma}$ 
& 4
& $(C,n,Y)$ \\
\hline
  \end{tabular}
  \caption{\label{classificationVSMbosons} 
  New vectors $X^\mu$ which can couple to $H$ or SM gauge bosons at the renormalizable level. 
  $(C, n, Y)$ denote generic quantum numbers under the SM gauge group. $R$ stands for a real $SU(3)_C$ representation (i.e.~$R = 1, 8, 27, \ldots$),  
  while $2k$ ($2k+1$) for an even (odd) $SU(2)_L$ representation. The subscript ``$\neq 1$'' means that the trivial representation is excluded.}
\end{table}

Note that some of the operators collected in \Table{classificationVSMbosons} can potentially 
yield extra non-standard contributions to the $g-2$. This happens for the operator 
$W_{\mu\nu} D^\mu X^\nu$, 
which only exists when $X$ transforms like $(1,3,0)$, or for some operators involving the $\epsilon$-tensor. 
It can be shown, however, that the former operator does not arise in renormalizable setups. 
To this end, let us consider the gauge embedding of the $SU(2)_L$ factor in terms of $SU(2)_1 \otimes SU(2)_2$ 
discussed in \sect{X130}: the only possible source of such operator is the kinetic term of the two field strengths 
\beq 
-\frac{1}{4} W^{a,\mu\nu}_{1} W^{a}_{1,\mu\nu}
-\frac{1}{4} W^{a,\mu\nu}_{2} W^{a}_{2,\mu\nu} \, ,
\eeq 
which upon an orthogonal transformation in terms of mass eigenstates, namely a massless triplet $W$ and a massive one $X$ 
(we neglect electroweak symmetry breaking here), leads to 
\beq 
-\frac{1}{4} W^{a,\mu\nu} W^{a}_{\mu\nu}
-\frac{1}{4} X^{a,\mu\nu} X^a_{\mu\nu} \, ,
\eeq 
without any $W$-$X$ mixed term. 
Similarly, among the operators obtained via an $\epsilon_{\mu\nu\rho\sigma}$ contraction, 
those arising from renormalizable theories are always total derivatives, and hence 
do not contribute to the $g-2$ in perturbation theory.

\vspace{1cm}

\bibliographystyle{utphys.bst}
\bibliography{bibliography}

\end{document}